\documentclass[letterpaper]{article}

\usepackage[letterpaper, margin=1in]{geometry}
\usepackage[T1]{fontenc}
\usepackage[utf8]{inputenc}
\usepackage{times}
\usepackage{microtype}
\usepackage[numbers,sort&compress]{natbib}

\usepackage{amsmath,amssymb,array,algorithm,algpseudocode}
\usepackage{url}
\usepackage{multirow}
\usepackage[table]{xcolor}
\usepackage{booktabs}
\usepackage{rotating}
\usepackage{subcaption}
\usepackage{graphicx}

\title{
Black-Box Adversarial Patch Attacks on VLAs \\via Ancestor VLM Exploitation
}

\author{
  \textbf{Xiaoyi Pang}\textsuperscript{1,*},
  \textbf{Haoyue Feng}\textsuperscript{2,*},
  \textbf{Quanxin Shou}\textsuperscript{1},\\
  \textbf{Yikun Miao}\textsuperscript{1},
  \textbf{Zhengyang Yan}\textsuperscript{1},
  \textbf{Song Guo}\textsuperscript{1} 
  \\[0.8em]
  \textsuperscript{1}The Hong Kong University of Science and Technology \quad
  \textsuperscript{2}Beihang University \\[0.5em]
  {\small \textsuperscript{*}Equal contribution 
  }
}

\date{}

\usepackage[hidelinks]{hyperref} 

\begin{document}

\maketitle

\begin{abstract}
Vision-Language-Action models (VLAs) are increasingly deployed in safety-critical physical environments, yet their adversarial robustness remains poorly understood. Existing attacks typically assume white-box access or rely on surrogate VLAs, which rarely holds in real-world deployments. 
Our key insight is that most VLAs are adapted from a publicly released pretrained vision-language model (VLM), inheriting two capabilities essential for action generation: visual perception and instruction-conditioned grounding.
Therefore, this paper explores a previously unaddressed question: can an adversary attack deployed VLAs using only their ancestor VLMs? To this end, we propose three adversarial patch attacks that disrupt the inherited capabilities: a vision disruption attack that corrupts the projected visual tokens through relative and absolute terms, an instruction-grounded semantic evidence suppression attack that removes the visual evidence required for instruction-grounded concepts, and a joint attack that unifies both objectives under a two-phase curriculum.
Experiments across different VLA families on both simulation and static real-world images show that patches optimized on the ancestor VLM cause substantial degradations in VLA task success rates, demonstrating that VLAs inherit adversarial vulnerabilities alongside their foundational capabilities. 
This effect is not uniform: it is strongest on tasks that require precise instruction-grounded localization, and nearly vanishes on policies whose adaptation rewrites the shared visual-semantic representation or whose action head iteratively smooths perturbations away.
By characterizing the boundary conditions of vulnerability inheritance and providing analysis of why the inheritance effect holds or fails, we advance the understanding of safety for VLA-involved systems.
\end{abstract}

\section{Introduction}
Vision-language-action models (VLAs) \citep{kim2024openvla, bu2025univla, black2024pi0, qu2025spatialvla, cen2025worldvla, pertsch2025fast} decide a robot's next action from an image of the scene and a natural-language instruction, combining visual perception, language understanding, and action generation in a single model. Most VLAs are not trained from scratch, but fine-tuned from a publicly-released pretrained vision-language model (VLM) \citep{shao2026largevlmbasedvisionlanguageactionmodels, 11164279, 11495231}. 
VLAs are increasingly deployed in physical environments such as industrial and scientific settings, homes, and UAVs \citep{zhu2026factoryfloordeploymentcasestudy, chen2025air, an2025measurement, lykov2025cognitivedrone}, where an adversarial attack \citep{9711337, Wang_2023_CVPR, Zheng_2025_ICCV, 10.1145/3658644.3670309, li2026visionlanguageactionsafetythreatschallenges} that can influence what the robot sees can make it fail a task and cause serious real-world consequences.
Understanding VLAs' adversarial robustness is therefore essential before these systems are deployed at scale.

Adversarial patch attacks place a small patch in the camera's view to corrupt what the model perceives and, through it, what it does, which is the most practical form of adversarial attack on robots. Prior patch attacks on VLAs fall into two groups by the access they assume. White-box attacks \citep{wang2025exploring, zhang2025attention, wang2025freezevla, huang2026trap, jones2025adversarial, yan2025alignment} assume full or partial access to the target VLA's parameters and action space. Transfer-based attacks \citep{xu2025model, lu2026robots, fu2026vla} require a surrogate VLA. Both assume more access than real deployments allow: the VLA is proprietary and no surrogate VLA is available. Attacks that require no access to the VLA's parameters or gradients therefore remain unexplored.

To close this gap, we aim to study an adversarial patch attack for VLAs that exploits only the publicly available ancestor VLMs from which the VLAs were fine-tuned. This is a practical black-box setting since the attacker never touches the VLA but exploits the fact that its foundation VLM is publicly accessible. If such an attack succeeds, it turns the common practice of fine-tuning public VLMs into a broad attack surface for VLAs. 
Our intuition is that a VLA inherits two capabilities that both are necessary for choosing the correct action from its ancestor VLM.
The first is visual perception, which builds a faithful description of what is in the scene, independent of the instruction. The second is instruction-conditioned grounding, which interprets the scene in light of the current instruction to identify what really matters, such as the target object, its distinguishing attributes, and its spatial relations.
Fine-tuning a VLM into a VLA retargets its output from text to actions, and robot fine-tuning data is far smaller and narrower than the VLM's pretraining corpus, so these shared representations largely survive fine-tuning.
This leads to a testable hypothesis: a patch that disrupts these capabilities in the ancestor VLM should also disrupt them, and hence the action decisions, in the derived VLA.

To test this hypothesis, we design three attack objectives on ancestor VLMs, one targeting each inherited capability and a third combining both. \textit{Vision Disruption} corrupts visual perception at the token level by measuring each adversarial token's drift from the clean token at its position along two axes: a relative term that scrambles the position-to-content mapping, and an absolute term that erases residual resemblance. \textit{Instruction-Grounded Semantic Evidence Suppression} uses the ancestor's own unembedding matrix as a logit lens to suppress the instruction-relevant evidence (i.e., object, attribute, relation) surviving in intermediate visual hidden states. \textit{Feature-and-Semantic Joint Disruption} combines both on one patch via a two-phase curriculum that activates the semantic loss only after the visual disruption stabilizes. We evaluate on VLAs fine-tuned from two VLM families, over simulated and real-world data, and confirm the hypothesis consistently.

In summary, our main contributions are as follows:
\begin{itemize}
    \item A new threat model: We identify and study the ancestor-only threat model for VLAs, and we provide the first systematic study of whether and when a VLM's adversarial vulnerability is inherited by its derived VLA.
    \item New attack methods: To achieve an adversarial patch attack on a VLA using only the publicly available ancestor VLM, we propose three attack objectives that target the two capabilities VLAs inherit from their ancestor VLMs, individually and jointly. 
    \item New findings: Through extensive experiments, we reveal that the public ancestor VLM is a shared supply-chain attack surface: a patch optimized on it degrades VLAs derived from it, across task domains, with no access to any target. We also characterize when this risk materializes: it peaks for the correct ancestor and for tasks requiring precise spatial localization, and recedes for fine-tuning that substantially shifts the shared representation away from the ancestor and for iterative flow-matching action heads, giving a principled rule for assessing which deployed VLAs are vulnerable.
\end{itemize} 

\section{Related work}
The adversarial attack usually misleads the model by adding carefully crafted perturbations to the input during inference. Existing adversarial attacks on VLAs can be broadly categorized by the degree of access assumed over the target model.

\textbf{White-box adversarial attacks on VLAs.} Wang et al. \citep{wang2025exploring} first conducted a systematic study on the adversarial robustness of VLA models, proposing three objectives for adversarial patches: UADA, which maximizes deviation in the predicted actions; UPA, which disrupts the spatial direction of end-effector movement; and TMA, which forces the VLA to predict specific trajectories. 
Then ADVLA \citep{zhang2025attention} adds attention-guided sparse image perturbations to destroy task execution, and FreezeVLA \citep{wang2025freezevla} uses bi-level optimization to induce action-freezing behavior. TRAP \citep{huang2026trap} optimizes a patch to hijack the intermediate reasoning and thus steer the chain-of-thought-based VLA toward adversary-chosen actions.
Beyond these visual-based attacks, Jones et al. \citep{jones2025adversarial} adapt LLM jailbreaking to VLA models by appending a short, optimized text suffix to the instruction. Yan et al. \citep{yan2025alignment} build VLA-Fool, a multimodal suite that combines textual, visual, and cross-modal perturbations.
While these methods achieve high attack success rates, they require access to the victim VLA's parameters and action space, rendering them inapplicable in a closed system.

\textbf{Transfer-based black-box adversarial attacks on VLAs.}
To relax the need for access to target VLA, several works have explored encoder-based attacks and transferable adversarial patches optimized on surrogate VLAs. 
EDPA \citep{xu2025model} disrupts image-text embedding alignment and maximizes visual feature divergence at the encoder level.
UPA-RFAS \citep{lu2026robots} jointly optimizes for visual-text misalignment, for dominating the model's attention, and for disturbing the predicted action on the surrogate VLA to achieve stronger cross-model transfer of the adversarial patch.
VLA-Hijack \citep{fu2026vla} identifies that all VLA-based robots must first use visual information to locate their own robotic arm before planning any motion, and exploits this shared visual self-localization process to achieve cross-VLA transferability.

All existing VLA adversarial attack methods require some victim-side or surrogate VLA access. We fill this gap by proposing the first attack framework that exploits only the ancestor VLM.

\section{Preliminary}

\begin{figure}[t]
    \centering
\includegraphics[width=0.98\linewidth]{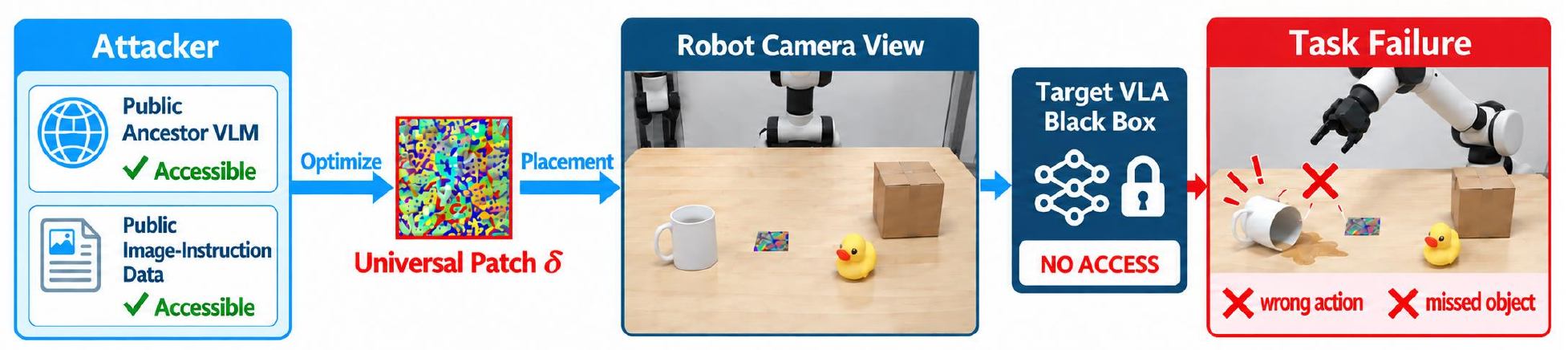}
    \caption{Threat Model}
    \label{fig:threat_model}
    \vspace{-3mm}
\end{figure}

\textbf{Threat Model.}
As shown in Figure. \ref{fig:threat_model}, we consider an attacker whose goal is to cause a deployed VLA to fail at its task by placing an adversarial patch at a random location in the robot's camera view.
The attacker has access to the public ancestor VLM from which the VLA was fine-tuned and a public image-instruction dataset representative of the VLA's tasks, but no access to the VLA's parameters, action space, or any surrogate VLA.
This is realistic because the ancestor VLM is usually publicly available.

\textbf{Attack Formulation.}
A VLA $\mathcal{F}_{\text{VLA}}$ maps an image $x$ and an instruction $t$ to an action $a$, i.e., $a = \mathcal{F}_{\text{VLA}}(x, t)$. We consider a single universal adversarial patch $\delta \in {[0,1]}^{h_p \times w_p \times C}$. 
At inference time, the patch is composited onto the observed image $x \in {[0,1]}^{H \times W \times C}$. Let $M$ be a binary mask that marks the patch region, the patched image is 
\begin{equation}
    x_{\text{adv}} = x \oplus \delta  = (1 - M) \odot x + M \odot \mathcal{T}(\delta),
\end{equation}
where the operator $\odot$ denotes Hadamard element-wise multiplication, and $\mathcal{T}$ is a random affine transformation that denotes geometric distortion due to viewing angle, distance, location, and surface variation. 
Let $a_{\text{adv}} = \mathcal{F}_{\text{VLA}}(x_{\text{adv}}, t)$ be the actions under the patched images, and $J$ be a task-failure loss that is large when the two actions differ. The ideal attack solves
\begin{equation}
   \delta^* = \arg\max \mathbb{E}_{(x,t)}\left[ J\big(\mathcal{F}_{\text{VLA}}(x_{\text{adv}}, t),\ \mathcal{F}_{\text{VLA}}(x, t)\big) \right]. 
\end{equation}
This objective requires querying the VLA, which our threat model does not allow. Our attacker therefore optimizes $\delta$ via a surrogate loss on its ancestor VLM $\mathcal{F}_{\text{VLM}}$.

\section{Methodology}
In this paper, we aim to explore whether adversarial patches optimized on the ancestor VLM will be effective on the derived VLA. 
Our intuition is: a VLA shares architecture and initialization with its ancestor, and its robot fine-tuning data is far smaller and narrower than the VLM's pretraining corpus, so the shared visual-semantic representation largely survives after fine-tuning. We therefore propose three optimization objectives to attack the two inherited capabilities that action generation depends on: visual perception and instruction-conditioned grounding.

\subsection{Objective 1: Vision Disruption}

A VLA choosing the correct action first requires visual feature recognition, which is the first capability the VLA inherits from its ancestor. Our first objective $\mathcal{L}_\text{VIS}$ therefore disrupts the projected visual tokens that the VLM consumes, so that the model no longer obtains a reliable description of the scene. We disrupt the post-projector tokens rather than the raw encoder features, because they are the exact visual interface the language decoder consumes (see Appendix \ref{app:post_projector}).

Prior work shows that measuring a patch's disruption at the token level is more damaging than measuring it at the image level: maximizing the mismatch between clean and adversarial visual tokens degrades a VLM far more than maximizing the same mismatch on the class token \citep{mei2026veattack}. Inspired by this, $\mathcal{L}_\text{VIS}$ breaks visual feature recognition at the projector output by measuring, from two complementary angles, how far each adversarial token has drifted from the clean token that occupied its position. 
Let $v^{\mathrm{clean}}_{b,i} \in \mathbb{R}^{D}$ denote the $i$-th projected visual token of the $b$-th clean image $x_b$, and $v^{\mathrm{adv}}_{b,i} \in \mathbb{R}^{D}$ the corresponding token of the patched image, both in the LLM embedding space after the projector. We $\ell_2$-normalize each
token and write $\hat v^{\mathrm{clean}}_{b,i}$, $\hat v^{\mathrm{adv}}_{b,i}$. Let $R$ be a cross-similarity matrix between adversarial and clean tokens within each sample, and we have $R_{b,i,j}=\hat v^{\mathrm{clean}\top}_{b,i}\hat v^{\mathrm{adv}}_{b,j}/\tau$.
The full loss is
\begin{equation}
\mathcal{L}_{\text{VIS}} = \underbrace{\frac{1}{B\cdot N_v}\sum_{b,i}\log\frac{\exp(R_{b,i,i})}{\sum_j\exp(R_{b,i,j})}}_{\mathcal{L}_\text{rel}^v}\ + \lambda_\text{vis-abs} \cdot\underbrace{\frac{1}{B\cdot N_v}\sum_{b,i} \hat{v}^{\text{clean}\top}_{b,i}\hat{v}^{\text{adv}}_{b,i}}_{\mathcal{L}_\text{abs}^v}.
\end{equation}
The relative term erases recognizability: it drives each adversarial token away from its own clean position, relative to every other position, so the position-to-content mapping is scrambled. Its softmax normalization, however, makes it insensitive to a uniform shift. The absolute term erases resemblance: it keeps reducing the raw per-position similarity even after the re-ranking saturates, thereby directly reducing residual same-position directional similarity rather than only changing the relative assignment among token positions. 
Both terms are computed over the full token grid rather than only the patch-covered cells since the patch covers only a small area of the image. Such grid-wide objectives therefore force the patch to propagate its corruption through self-attention to non-covered locations.

\subsection{Objective 2: Instruction-Grounded Semantic Evidence Suppression}\label{sec:objective2}
A VLA choosing the correct action also requires grounding the scene in light of the instruction, e.g., identifying the target object, the attributes that distinguish it from similar objects, and the relation that locates it. We refer to concepts such as object, attribute, and relation as the instruction-relevant concepts.
We hypothesize the instruction-conditioned grounding rests on visual evidence for these concepts, which has not yet formed at the projector output but emerges in the visual-token states at the intermediate layers of the backbone. Our second objective $\mathcal{L}_\text{IGS}$ therefore suppresses that evidence in visual-token states of the ancestor VLM. The instruction enters the objective only through the probe set, which selects the concepts to erase.

\textbf{Logit-lens evidence scoring under instruction-grounded semantic probe.} For each image-instruction pair we precompute factorized side labels, i.e., object, attribute, and relation, obtained offline either by parsing the instruction directly or by prompting the ancestor VLM for a description and factorizing it (see Appendix \ref{app:probe and evidence}). They form the probe set $\mathcal{Q}_b$, and each probe token $q$ is an instruction-grounded semantic concept whose visual evidence we aim to suppress. 
We reuse the ancestor's unembedding matrix $W_{\mathrm{lm}}$ as a \emph{logit lens}: projecting an intermediate state into vocabulary space reveals how strongly it aligns with a word, without running the remaining decoder layers. Let $h^{(l)}_{b,i}$ be the hidden state of visual token $i$ at layer $l$ and $\{w_{q,k}\}_{k=1}^{N_w}$ denote the rows of $W_{\mathrm{lm}}$ for the subword tokens of $q$. The evidence for a probe $q$ at visual token $i$ of layer $l$ is the inner product between the hidden state $h^{(l)}_{b, i}$ after the final RMSNorm and the subword vectors of $q$ from the LLM's output matrix, averaged over subwords and aggregated over probes:
\begin{equation}
s_{b,i}(q)=\frac{1}{N_w}\sum_{k=1}^{N_w}\text{RMSNorm}(
h^{(l)}_{b,i})^{\top}\,w_{q,k}, \quad
s_{b,i}=\frac{1}{|\mathcal{Q}_b|}\sum_{q\in \mathcal{Q}_b}s_{b,i}(q),
\end{equation}
This is a representation-level alignment score: how strongly does the hidden state at this visual token position point in the direction that the model's own representation space associates with the concept $q$? A high score means the hidden state is geometrically encoding $q$-like content in the model's internal language.
\textbf{Clean-derived importance weighting}. Since evidence is not spread uniformly across tokens and much of the scene is background that the instruction does not care about, spending the disruption budget on those tokens would be wasteful. We therefore aggregate over tokens with weights fixed from the clean image:
\begin{equation}
\alpha_{b,i} = \frac{\exp(s_{b,i}^{\text{clean}}/ \tau_{\text{sel}})}{\sum_{j}\exp(s_{b,j}^{\text{clean}}/ \tau_{\text{sel}})},
\qquad
    S_{b}^{\text{clean}} =
\sum_{i}\alpha_{b,i}\cdot s_{b,i}^{\text{clean}},
\qquad
S_{b}^{\text{adv}} =
\sum_{i}\alpha_{b,i}\cdot s_{b,i}^{\text{adv}}.
\end{equation}

\textbf{Two-term loss.} We express the required suppression as two constraints, including both the relative and absolute ones. Specifically, the relative term pushes the adversarial evidence below its clean value, and the absolute term pushes it below a target $S_{tgt}$ (an upper bound):
\begin{equation}\label{eq:IGS}
\mathcal{L}_{\mathrm{IGS}}=\underbrace{\frac{1}{B}\sum_b
\mathrm{softplus}\!\left(\frac{S^{\mathrm{adv}}_b-S^{\mathrm{clean}}_b+m}{\tau_{\text{drop}}}\right)}_{\mathcal{L}_{\mathrm{rel}}^s}
+\lambda_\text{sem-abs}\underbrace{\frac{1}{B}\sum_b
\mathrm{softplus}\!\left(\frac{S^{\mathrm{adv}}_b-S_{\mathrm{tgt}}}{\tau_{\text{drop}}}\right)}_{\mathcal{L}_{\mathrm{abs}}^s},
\end{equation}

where $S_{\mathrm{tgt}}$ is set as the $\rho$-th percentile of the clean evidence distribution $\{S^{\mathrm{clean}}_b\}$. We set $\rho=10$ as the default.
The softplus keeps the gradient non-zero below the boundary, so patch keeps pushing instead of stalling once the constraint is met.

\subsection{Objective 3: Feature and semantic joint disruption}
Each objective above disrupts a different aspect of the ancestor-VLM representation, and their combination is intended to improve transfer robustness. $\mathcal{L}_\text{VIS}$ only corrupts the projected visual representation, but it leaves the instruction channel intact, so the model can still use the instruction as a semantic prior to partly recover the scene. $\mathcal{L}_\text{IGS}$  targets an instruction-relevant semantic readout from intermediate visual-token states. The instruction guides which concept directions are probed, rather than directly conditioning the probed visual-token states.
We therefore optimize both objectives on a single patch.
Combining them, however, is not as simple as summing two losses. The two objectives are pointed at different stages of the pipeline, and we observe that the pixel-space directions that disrupt projector-level visual tokens are largely independent of the directions that suppress intermediate-layer semantic evidence.

To be specific, we adopt a two-phase curriculum. In Phase~1 (steps $1,\dots,K_1$) we optimize $\mathcal{L}_{\text{VIS}}$ alone, which lets the patch settle into a stable direction in pixel space. In
Phase~2 (steps $K_1+1,\dots,K$),  $\mathcal{L}_{\text{IGS}}$ becomes eligible for activation, subject to the visual-basin gate: it is admitted only when the exponential moving average of the global cosine similarity $c_t$ between clean and adversarial visual tokens has dropped below a threshold $\gamma$, ensuring that the $\mathcal{L}_{\text{IGS}}$ gradient cannot undo that visual disruption (see Appendix \ref{app:staged}). Writing $\beta_k=\mathbb{I}(k > K_1)\cdot\mathbb{I}(c_t\le \gamma)$, 
the objective can be expressed as:
\begin{equation}
\mathcal{L}_{\mathrm{joint}}
=\mathcal{L}_{\mathrm{VIS}}
+\beta_k\,\lambda_{\mathrm{sem}}\,\mathcal{L}_{\mathrm{IGS}}.
\end{equation}

\section{Evaluation}

\subsection{Experimental Setup}
\textbf{Datasets and Models}. 
We optimize patches on multiple datasets in separate experiments. LIBERO \citep{liu2023libero} is a simulation benchmark with four suites, Spatial, Object, Goal, and Long, each containing 10 tasks evaluated over 50 rollouts. BridgeData V2 \citep{walke2023bridgedata} is a large real-world dataset with 60,096 trajectories.
We evaluate our attack on three victim VLAs: OpenVLA \citep{kim2024openvla}, UniVLA \citep{bu2025univla}, and $\pi_{0.5}$ \citep{intelligence2025pi0.5}. OpenVLA and UniVLA are adapted from Prismatic-7B \citep{karamcheti2024prismatic}, and $\pi_{0.5}$ from PaliGemma-3B \citep{beyer2024paligemma}. 
All of these victim models provide fine-tuned variants for LIBERO. 
Specifically, OpenVLA offers separate fine-tuned models for individual task suites, whereas both UniVLA and $\pi_{0.5}$ are fine-tuned as a single unified model across all task suites.

\textbf{Baselines}. Our threat model, in which a patch is optimized on the ancestor VLM alone to attack a VLA, has no directly comparable prior method. We therefore consider two naive baselines, as well as adapting two state-of-the-art transfer-based attacks to our settings. 
1) Random: random noise in $[0,1]$ of identical size and position. 2) Blank: uniform mid-grey patch, isolating occlusion. 3) EDPA \citep{xu2025model}, adapted to the ancestor: its losses are computed on the ancestor instead of the victim's encoder. 4) UPA-RFAS \citep{lu2026robots}, adapted to the ancestor with its feature-space repulsive InfoNCE objective and $l_1$ deviation loss.

\textbf{Evaluation Metrics}. We report task success rate, where a lower value means a stronger attack performance. We optimize five patches with different random seeds and report the mean.

\textbf{Implementation \& Evaluation Details}. 
The patch occupies only $5\%$ of the image input, and is placed at a uniformly random location both during
optimization and at test time.
We optimize it with Adam ($\beta=(0.9,0.999)$) for 3500 outer steps. Each outer step samples a mini-batch of $B=4$ frames and performs 4 Adam updates on it. Each update averages gradients over 8 EoT draws (including geometric: random shear and rotation; and photometric: random brightness, contrast, and Gaussian noise).
We probe the output of the zero-indexed Llama decoder layer 12, retain only the image-token states, apply the frozen final RMSNorm, and compute logit-lens scores using selected rows of the frozen language-model head. We set $\lambda_{\mathrm{sem}}=0.5$, $\lambda_\text{vis-abs}=2.0$, and $\lambda_\text{sem-abs}=1.0$.
Since $\pi_{0.5}$ consumes two views (agentview and eye-in-hand), we train a primary patch on the agent view and a wrist patch on the eye-in-hand view, applied at independent random locations. More details are provided in Appendix \ref{app:experiment-details}.


\subsection{Attacking VLAs via the Ancestor VLM: Main Results}
We optimize patches on Prismatic-7B and evaluate on OpenVLA and UniVLA using the LIBERO benchmark. Patches are separately optimized with LIBERO data and BridgeData V2, covering two scenarios where the attacker knows or doesn't know the target domain. Table. \ref{table: openvla-class} reports task success rates across four task suites.

\begin{table}[t]
\centering
\caption{Task success rate (\%) on LIBERO for OpenVLA and UniVLA (lower is a stronger attack). \textbf{No Attack}/\textbf{Random}/\textbf{Blank} are patch-source-independent and listed once. The remaining attacks are grouped by whether the adversarial patch was optimized on the ancestor Prismatic-7B using LIBERO or BridgeData V2.}
\label{table: openvla-class}
\small
\setlength{\tabcolsep}{3.5pt}
\renewcommand{\arraystretch}{1.1}
\begin{tabular}{l|ccccc|ccccc}
\toprule
\multirow{2}{*}{\textbf{Attack}} & \multicolumn{5}{c|}{\textbf{OpenVLA}} & \multicolumn{5}{c}{\textbf{UniVLA}} \\
 & \textbf{Spatial} & \textbf{Object} & \textbf{Goal} & \textbf{Long} & \textbf{Avg} & \textbf{Spatial} & \textbf{Object} & \textbf{Goal} & \textbf{Long} & \textbf{Avg} \\
\hline
\textbf{No Attack} & 79.4 & 68.8 & 77.4 & 49.4 & 68.75 & 88.8 & 98.1 & 96.8 & 88.6 & 93.08 \\
\textbf{Random}    & 66.2 & 48.0 & 60.2 & 21.5 & 48.98 & 71.6 & 64.0 & 80.4 & 54.2 & 67.55 \\
\textbf{Blank}     & 62.0 & 49.4 & 57.6 & 23.6 & 48.15 & 77.6 & 75.6 & 84.2 & 60.0 & 74.35 \\
\hline
\multicolumn{11}{c}{\textit{Patch optimized with LIBERO}} \\
\hline
\textbf{EDPA}       & 58.0 & 47.6 & 57.0 & 22.2 & 46.20 & 61.6 & 57.4 & 75.8 & 41.6 & 59.10 \\
\textbf{UPA-RFAS}   & 55.0 & 39.2 & 53.6 & 18.2 & 41.50 & 60.0 & 54.8 & 72.8 & 35.8 & 55.85 \\
\textbf{Ours-VIS}   & 42.0 & 35.4 & 47.0 & 17.6 & 35.50 & 43.8 & 52.6 & 70.0 & 25.8 & 48.05 \\
\textbf{Ours-IGS}   & 48.0 & 46.2 & 54.2 & 17.2 & 41.40 & 58.2 & 54.8 & 75.6 & 37.4 & 56.50 \\
\textbf{Ours-Joint} & 39.0 & 34.2 & 46.4 & 13.2 & 33.20 & 42.0 & 50.2 & 68.0 & 23.6 & 45.95 \\
\hline
\multicolumn{11}{c}{\textit{Patch optimized with BridgeData V2}} \\
\hline
\textbf{EDPA}       & 56.6 & 45.6 & 57.6 & 18.2 & 44.50 & 65.0 & 57.2 & 75.2 & 42.6 & 60.00 \\
\textbf{UPA-RFAS}   & 54.4 & 45.6 & 56.2 & 17.0 & 43.30 & 60.4 & 57.8 & 73.4 & 35.0 & 56.65 \\
\textbf{Ours-VIS}   & 44.0 & 35.6 & 44.6 & 16.8 & 35.25 & 47.8 & 53.8 & 73.6 & 31.6 & 51.70 \\
\textbf{Ours-IGS}   & 45.8 & 39.8 & 47.6 & 14.5 & 36.92 & 49.2 & 52.2 & 73.0 & 30.0 & 51.10 \\
\textbf{Ours-Joint} & 40.6 & 34.4 & 44.6 & 12.0 & 32.90 & 43.8 & 51.2 & 69.2 & 23.2 & 46.85 \\
\bottomrule
\end{tabular}
\vspace{-3mm}
\end{table}

\textbf{Our attacks are effective}. All three of our attacks reduce task success rates well below the random patch and blank patch baselines on both OpenVLA and UniVLA, and far below the no-attack level. Since our adversarial patches are optimized entirely on the ancestor VLM, this is direct evidence that adversarial vulnerability is inherited from the VLM to its derived VLA.

\textbf{Our attacks outperform the adapted transfer-based baselines}.
Ours-VIS reaches 35.5\% average SR on OpenVLA and 48.05\% on UniVLA, beating both EDPA (46.2\% and 59.1\%) and UPA-RFAS (41.5\% and 55.85\%), and Ours-joint is stronger still. In the ancestor-only setting, both EDPA and UPA-RFAS reduce to a weaker objective than our two-term objective. 
To be specific, EDPA combines a patch contrastive loss that scrambles the per-token position-to-content mapping and an image-instruction alignment loss that pushes the global visual embedding away from the paired instruction embedding. 
On robot data, however, the VLM's image and instruction embeddings are already poorly aligned, since the VLM was not pretrained on such data, so EDPA reduces to its patch-contrastive term, which acts on the encoder output before the projector.
UPA-RFAS combines a feature-space objective, an $\ell_1$ deviation prior plus a repulsive InfoNCE that pushes patched features away from their clean anchors, with two VLA-specific attention and semantic losses. With those VLA-specific losses disabled in the ancestor VLM, the surviving feature-space repulsion lacks our per-token relative-and-absolute disruption at the projector output.

\textbf{In most cases, Ours-IGS alone is weaker than Ours-VIS alone, and Ours-joint outperforms either objective alone. }
For example, on OpenVLA, when patches are optimized with LIBERO, Ours-IGS reduces the average success rate to only 41.4\%, a weaker effect than Ours-VIS (35.5\%); when patches are optimized with BridgeData V2, Ours-IGS and Ours-VIS achieve average success rates of 36.92\% and 35.25\%, respectively.
This is because  $\mathcal{L}_{\mathrm{IGS}}$ attacks deep intermediate states where the gradient path to the patch is long and noisy. Its value emerges in combination with $\mathcal{L}_{\mathrm{VIS}}$: Ours-joint is the strongest in every suite on both models (33.2\% vs. 35.5\% on OpenVLA, and 45.95\% vs. 48.05\% on UniVLA).
The reason is that the semantic term $\mathcal{L}_{\mathrm{IGS}}$ targets the task-relevant evidence that survives visual disruption, so the two objectives together remove both the perceptual signal and the model's ability to compensate through instruction-guided attention.

\textbf{Cross-Domain Transfer}. 
Optimizing on LIBERO versus BridgeData V2 gives similar attack performance on the same LIBERO episodes. Measured as $BridgeData - LIBERO$, the differences are $-0.25/+3.65$ for Ours-VIS and $-0.3/+0.9$ for Ours-joint (OpenVLA/UniVLA). This means an attacker does not need to know the target's task domain and can transfer across domains with little loss, which further lowers the bar for a practical attack. 
The exception is Ours-IGS, which is consistently stronger with BridgeData V2. To rule out a data-size effect, we re-optimized it on a BridgeData V2 subset of the same size as the LIBERO set. On this subset, it reaches 39.2\% on OpenVLA and 53.4\% on UniVLA, lower than the 41.4\% and 56.5\% obtained when optimizing on LIBERO. 
Meanwhile, as we can observe in the table, the full BridgeData V2 set lowers the rates further, to 36.9\% and 51.1\%. 
The reason is: $\mathcal{L}_{\mathrm{IGS}}$ learns which concepts to suppress from the training frames. A single LIBERO suite exposes it to only a few objects and relations, whereas real-world BridgeData V2 frames span many scenes and tasks. Such a dataset can help the loss learn the general rule, so as to suppress whatever the instruction refers to, and that rule carries over to suites the patch was never optimized on.


\textbf{Task-dependent attack performance}. To isolate the specific effect of our patch from each suite's baseline fragility, we measure the success-rate drop relative to a random patch. On OpenVLA, this advantage is largest on the Spatial suite, where Ours-joint reaches 39.0\% against 66.2\% for a random patch (a 27.2 pp advantage), and smallest on the Long suite (13.2\% against 21.5\%, an 8.3 pp advantage). Spatial tasks hinge on the precise instruction-grounded object localization that our objectives target, whereas long-horizon tasks fail for many reasons beyond single-frame perception, so a perception attack adds relatively little on top of a random patch.

\begin{figure}[t]
  \centering
  \begin{minipage}[t]{0.27\textwidth}
    \centering
    \includegraphics[width=\textwidth]{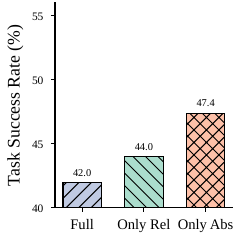}
\caption{$\mathcal{L}_{\mathrm{VIS}}$ Ablation.}
    \label{fig:L_vis ablation}
  \end{minipage}
  \hfill 
  \begin{minipage}[t]{0.27\textwidth}
    \centering
    \includegraphics[width=\textwidth]{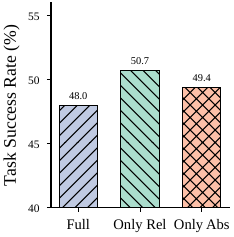}
    \caption{$\mathcal{L}_{\mathrm{IGS}}$ Ablation.}
    \label{fig:L_IGS ablation}
  \end{minipage}
  \hfill
  \begin{minipage}[t]{0.4\textwidth}
    \centering
    \includegraphics[width=\textwidth]{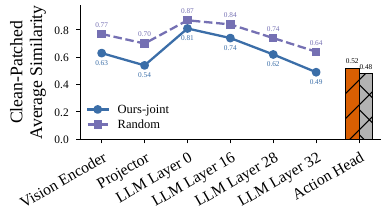}
    \caption{Attack verification.}
    \label{fig:3}
  \end{minipage}
  \vspace{-5mm}
\end{figure}

\subsection{Ablation Study}

\textbf{Contribution of each loss term}.
Figure \ref{fig:L_vis ablation} and \ref{fig:L_IGS ablation} show the effect of the components of $\mathcal{L}_{\mathrm{VIS}}$ and $\mathcal{L}_{\mathrm{IGS}}$, and the results are tested on OpenVLA with the LIBERO-spatial suite. 
We can find that each term alone already reduces the success rate, but the combination is stronger. 
This confirms that the two terms we design are complementary rather than redundant.


\textbf{Mechanistic verification}. To trace where and how our attack achieves its intended effect, we feed the clean and patched versions of the same frame through the victim VLA and compare two things: the hidden states of the visual tokens at different layers, and the decoded actions. To be specific, we compute the mean cosine similarity between clean and patched visual-token hidden states. For actions, we compute the average $\ell_2$ distance $d$ between the decoded actions under clean and patched observations, and report the similarity $s_{\mathrm{act}} = e^{-d} \in (0,1]$. The results are shown in Figure \ref{fig:3}. All quantities are averaged over the OpenVLA-LIBERO-Spatial suite, using the patch optimized on the ancestor Prismatic-7B with $\mathcal{L}_{\mathrm{joint}}$. 
We can find that at every stage we measure, our attack consistently leads to lower clean-patched similarity than the random patch attack, indicating our stronger attack performance.
Besides, the mean cosine similarity between hidden states first rises from the projector output to the first LLM layer, where self-attention dilutes the patch's local corruption across the mostly clean visual tokens, and then falls through the deeper layers as the corruption diffuses. This early partial recovery is why $\mathcal{L}_{\mathrm{VIS}}$ alone is insufficient. $\mathcal{L}_{\mathrm{IGS}}$ applied at a deeper layer removes the residual evidence that survives it.
Correspondingly, the action similarity is also lower than for the random
patch (0.48 vs. 0.52), showing that this representation disruption propagates into the VLA's
action choices.

\begin{table}[t]
\centering
\small 
\renewcommand{\arraystretch}{1.1}

\begin{minipage}[t]{0.45\textwidth}
    \centering
    \setlength{\tabcolsep}{4pt}
    \caption{Task success rate (\%) on OpenVLA across LIBERO suites, when the patches are optimized on different VLMs using LIBERO data.}
    \label{tab:different ancestor}
    \begin{tabular}{l ccccc}
    \toprule
    \textbf{Attack} & \textbf{Spa.} & \textbf{Obj.} & \textbf{Goa.} & \textbf{Lon.} & \textbf{Avg} \\
    \midrule
    No Attack & 79.4& 68.8& 77.4& 49.4& 68.75\\
    \midrule
    Random  & 66.2& 48.0& 60.2& 21.5& 48.98\\
    Blank& 62.0& 49.4& 57.6& 23.6& 48.15\\
    \midrule
    
    \rowcolor{gray!10} \multicolumn{6}{l}{\textbf{Patch optimized on Prismatic-7B}} \\
    Ours-VIS   & 42.0& 35.4& 47.0& 17.6& 35.5\\
    Ours-IGS   & 48.0& 46.2& 54.2& 17.2& 41.4\\
    Ours-joint & 39.0& 34.2& 46.4& 13.2& 33.2\\
    \midrule
    \rowcolor{gray!10} \multicolumn{6}{l}{\textbf{Patch optimized on LLaVA-1.5 7B}} \\
    Ours-VIS   & 56.8&	45.9&	57&	18.6&	44.58\\
    Ours-IGS   & 57.4 &48.8&	56&	17.4& 44.9	\\
    Ours-joint & 54.8&	44.5&	56.8&	19&	43.78\\
    \midrule
    \rowcolor{gray!10} \multicolumn{6}{l}{\textbf{Patch optimized on PaliGemma-3B}} \\
    Ours-VIS   & 63.8&	48.3&	62.6&	21.4&	49.03\\
    Ours-IGS   & 62.4&	50.4&	64&	22.4&	49.8\\
    Ours-joint & 62& 49.9& 61&	22&	48.73\\
    \bottomrule
    \end{tabular}
\end{minipage}
\hfill 
\begin{minipage}[t]{0.48\textwidth}
    \centering
    \setlength{\tabcolsep}{5pt}
    \caption{Task success rate (\%) on $\pi_{0.5}$ under LIBERO. Patches are optimized on PaliGemma-3B, except UADA, a white-box baseline optimized on OpenVLA.}
    \label{tab:pi0}
    \begin{tabular}{l ccccc}
    \toprule
    \textbf{Attack} & \textbf{Spa.} & \textbf{Obj.} & \textbf{Goa.} & \textbf{Lon.} & \textbf{Avg} \\
    \midrule
    No Attack & 98.4 & 98.2 & 97.2 & 92.6 & 96.6 \\
    \midrule
    Random    & 95.6 & 98.2 & 95.6 & 89.0 & 94.6\\
    Blank     & 95.8 & 98.0 & 95.8 & 88.0 & 94.4\\
    \midrule
    \rowcolor{gray!10} \multicolumn{6}{l}{\textbf{Patch optimized with LIBERO}} \\
    UADA& 93.2 & 93.4 & 92.4 & 71.6 & 87.7\\
    UPA-RFAS   & 95.8&	96.8&	92.4&	86.4&	92.85\\
    Ours-VIS   & 95.2&	96.8&	96.8&	85.8&	93.65\\
    Ours-IGS   & 95.8&	96.6&	94.4&	82&	92.2\\
    Ours-joint & 95.8&	97.2&	94.8&	86.6&	93.6\\
    \addlinespace
    \midrule
    \rowcolor{gray!10} \multicolumn{6}{l}{\textbf{Patch optimized with BridgeData V2}} \\
    UADA& 90.6 & 83.2 & 86.4 & 71.4 & 82.9\\
    UPA-RFAS   & 96.0 & 98.2 & 95.8 & 86.4 & 94.1\\
    Ours-VIS   & 94&	98.2&	96.2&	87.2&	93.9\\
    Ours-IGS   & 95.8&	96.6&	95.2&	84.4&	93\\
    Ours-joint & 95.2&	97.4&	95.6&	84.6&	93.2\\
    \addlinespace
    \bottomrule
    \end{tabular}
\end{minipage}

\end{table}

\subsection{What Determines Cross-Architecture Transfer?}
We investigate how the degree of architectural similarity between the surrogate VLM and the target VLA affects attack transferability. We evaluate our attacks optimized on three surrogate models against OpenVLA: Prismatic-7B (direct ancestor), LLaVA-1.5-7B (shares the Llama-2 backbone but uses a different visual encoder — CLIP ViT-L/14 rather than SigLIP + DINOv2), and PaliGemma-3B (a different language backbone and a SigLIP-based encoder rather than the ancestor's fused DINOv2 + SigLIP backbone). The results are shown in Table \ref{tab:different ancestor}.

We can observe that the attack transferability degrades with architectural distance as the surrogate model diverges from the target VLA's ancestry: Prismatic-7B (direct ancestor) achieves the strongest attack (Ours-VIS, Ours-IGS and Ours-Joint reach 35.5\%, 41.4\% and 33.2\% average success rates), LLaVA-1.5-7B (partial overlap) produces intermediate results (our three objectives reach 44.58\%, 44.9\% and 43.78\%), and PaliGemma-3B (no shared language backbone and a different visual backbone) yields performance close to the random patch baseline. The same ordering appears suite by suite. The comparison between Prismatic-7B and LLaVA-1.5-7B is particularly informative: both share the Llama-2-7B language backbone with OpenVLA, but only Prismatic-7B shares the visual encoder, and Prismatic-7B transfers substantially better, showing that sharing the LLM backbone alone is not sufficient for strong transfer.

\subsection{When Does the Attack Fail? Boundary Conditions}
We test the attack on $\pi_{0.5}$, which is fine-tuned from PaliGemma-3B under a substantially different training regime: a large fraction of PaliGemma's 3B parameters are updated during $\pi_{0.5}$ training, and actions are generated through a separate flow-matching expert that iteratively denoises a noisy action trajectory over its denoising steps. The results are shown in Table \ref{tab:pi0}.

We can observe that on $\pi_{0.5}$, our attacks and adapted transfer-based baselines all collapse to near-random success rates, and even the white-box, action-based attack UADA optimized on OpenVLA transfers only weakly (87.7\% with LIBERO and 82.9\% with BridgeData V2).
$\pi_{0.5}$ is therefore fundamentally hard to attack through visual-patch transfer. Two architectural properties account for this. First, PaliGemma's large-scale parameter update during $\pi_{0.5}$ training substantially shifts the internal representations away from the original VLM distribution, reducing the alignment between the surrogate model (original PaliGemma) and the target ($\pi_{0.5}$). Second, the flow-matching action expert performs many steps of iterative denoising, which smooths out upstream perturbations before they reach the action output. 
These results identify two conditions under which the attack fails: fine-tuning that substantially rewrites the shared representation away from the ancestor, and an iterative action head that smooths the perturbation before it reaches the action output. 

\section{Conclusion and Discussion}
In this paper, we showed that an adversary can attack a deployed VLA using only its public ancestor VLM, turning the common practice of fine-tuning public VLMs into a shared, supply-chain attack surface. This risk is not uniform: it peaks when the VLA preserves the ancestor's visual-semantic representation and reads it directly for action generation, and recedes when fine-tuning substantially rewrites that representation or an iterative action head smooths the perturbation away, giving practitioners a concrete rule for assessing which deployed VLAs are vulnerable.

Our attack has several limitations.
First, it is weak against $\pi_{0.5}$-like architectures, which smooth out upstream perturbations.
Second, it depends on using the correct ancestor, and its strength degrades with architectural distance.
Third, our evaluation is on simulation and static real-world images, not physical robots.
Finally, our threat model assumes the attacker knows the ancestor VLM; if this information is concealed, the attack surface shrinks considerably.

We have implications for defense. First, defense should be organized around the ancestor rather than only the individual policy. Since one patch transfers to multiple VLAs derived from the same ancestor, a single patch bank per ancestor can serve as a shared red-team artifact, and the same patches can be reused for adversarial training. Second, our attacks are invisible to query-based monitoring, since the adversary never touches the victim, so detection has to come from the input side or from the robot itself. Our own objectives point to the cues: the vision loss maximizes per-token drift away from the clean image, so a token-level drift tripwire is a natural monitor, and the semantic loss removes object, attribute, and relation evidence, so a grounding-consistency check against an independent detector may catch the failures that matter most. We do not evaluate any of these defenses ourselves, but our results turn them into concrete, testable predictions.

\bibliography{reference}
\bibliographystyle{plainnat}

\clearpage
\appendix

\section{More Details about Experiments}
\label{app:experiment-details}

\subsection{Patch composition}
\label{app:patch-paste}

\begin{figure}[t]
	\centering
	\includegraphics[width=\linewidth]
	{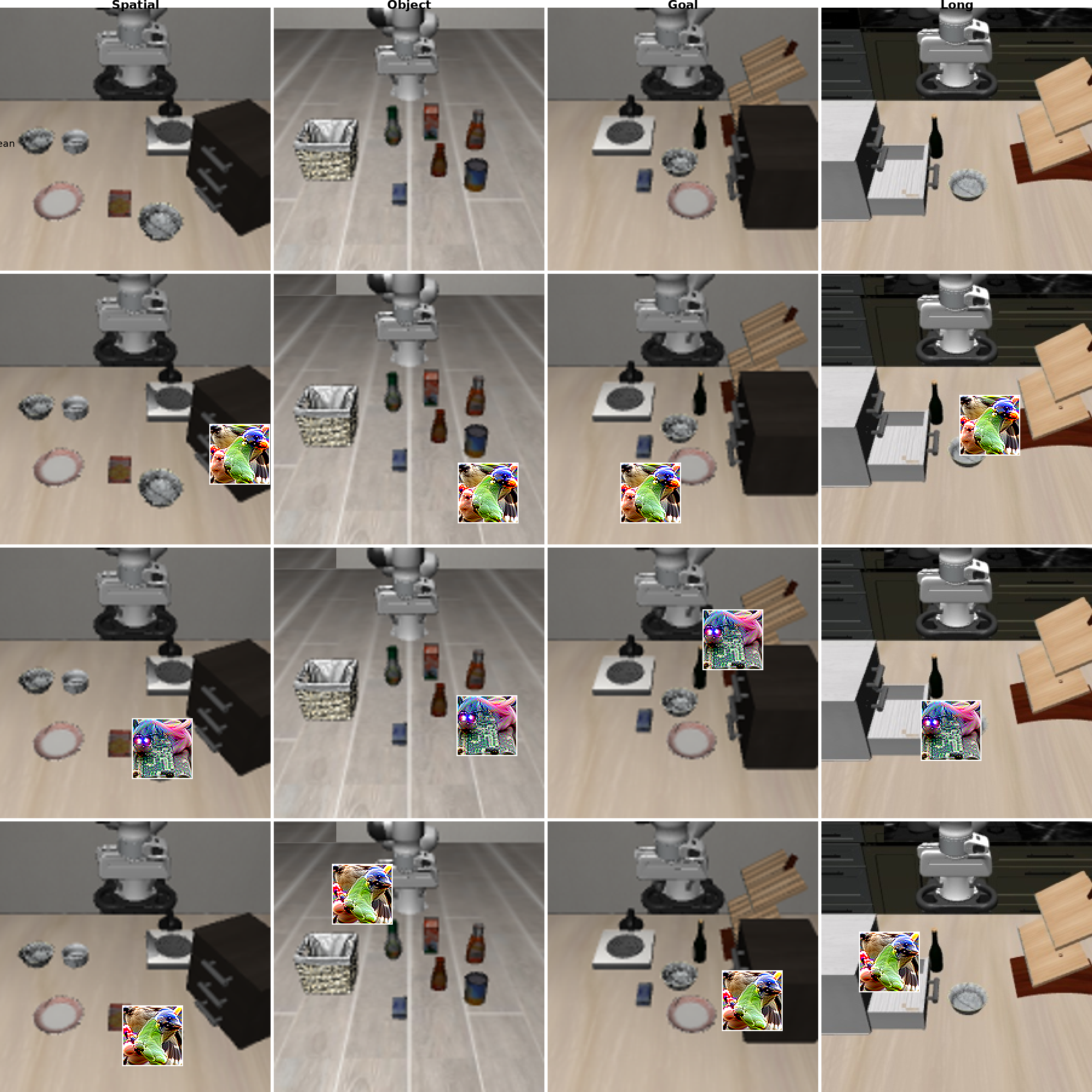}
	\caption{Random-location overlays of the patches optimized on $\mathcal{L}_\text{VIS}$, $\mathcal{L}_\text{IGS}$, and $\mathcal{L}_\text{Joint}$ on LIBERO observations
		(columns: Spatial, Object, Goal, Long; rows: clean, $\mathcal{L}_\text{VIS}$, $\mathcal{L}_\text{IGS}$, $\mathcal{L}_\text{Joint}$).}
	\label{fig:patch-variants-random}
\end{figure}

We optimize a square RGB patch $\delta \in[0,1]^{3\times s\times s}$ on the ancestor VLM's $224\times224$ canvas.  With area fraction $0.05$, $s=\mathrm{round}(224\sqrt{0.05})=50$.
The patch is composited by hard overwrite through an affine warp so that only covered pixels are replaced, with gradients flowing back through the warp. 
Figure~\ref{fig:patch-variants-random} shows example overlays.


\subsection{Training placement, EOT, and gradient updates}
\label{app:train-eot}


\textbf{Loop structure.} Our optimizer has three levels.
\begin{itemize}
  \item $T=3{,}500$ \emph{optimization iterations}. The learning-rate schedule (cosine decay
        after a $100$-iteration linear warm-up) is defined over these iterations.
  \item $4$ \emph{Adam updates} per iteration, i.e.\ $14{,}000$ in total
        ($\text{lr}=10^{-3}$, $\beta=(0.9,0.999)$).
  \item $8$ \emph{expectation-over-transformation (EoT) draws} per update, whose gradients are
        averaged before the update is applied.
\end{itemize}

\textbf{Placement.} Each EoT draw samples one patch location uniformly over the canvas and shares it across the batch of $4$; the location is held fixed within a draw and resampled at the next one. This mirrors the deployment assumption that a physical sticker occupies one pose per observation burst while still randomizing pose throughout optimization. 

\textbf{EOT draws.}
Each inner update accumulates gradients over 8 independent EOT draws. A single draw applies, in order:
\begin{enumerate}
	\item uniform random placement of the $50\times50$ patch on the $224$ canvas;
	\item with probability $0.8$, a geometric warp of rotation 	$\mathrm{Unif}[-30^\circ,30^\circ]$ and shear $\mathrm{Unif}[-0.2,0.2]$
	(identity with probability $0.2$);
	\item photometric jitter on the \emph{composited} frame: brightness and
	contrast factors $1\pm 0.05$, plus i.i.d.\ Gaussian noise of std $0.02$,
	then clamp to $[0,1]$.
\end{enumerate}

\textbf{Ancestor- and view-specific training.}
Every patch is optimized against the frozen ancestor of the victim it is intended for; no victim parameters are involved at any point.
\begin{itemize}
  \item \emph{Prismatic-7B (OpenVLA, UniVLA).} One patch, trained on agent-view frames of a $224\times224$ canvas. The two vision towers are fused inside the frozen ancestor, so the patch has a single parameterization.
  \item \emph{PaliGemma-3B ($\pi_{0.5}$).} $\pi_{0.5}$ consumes an agent view and an eye-in-hand stream, so we train two patches with the same objective on disjoint data: a primary patch on agent-view frames and a wrist patch on eye-in-hand frames. They are never optimized jointly. At evaluation the primary patch is composited on the agent view and the wrist patch on the eye-in-hand stream, at independently drawn locations.
  \item \emph{LLaVA-1.5-7B (surrogate control).} The patch is composited on the $224$ canvas and the image is then resized to LLaVA's native $336$. Training follows the OpenVLA single-view setting.
\end{itemize}

\subsection{Evaluation protocol}
\label{app:eval-details}
\paragraph{Common LIBERO settings.}
Unless stated otherwise, every reported success rate uses $10$ tasks $\times$ $50$ trials $=500$ episodes per suite (\texttt{libero\_spatial}, \texttt{libero\_object}, \texttt{libero\_goal},\texttt{libero\_long}), wait steps $10$ before control, and suite-wise horizon caps of $220$ / $280$ / $300$ / $520$ environment steps (Spatial / Object / Goal / Long).  Patches cover $5\%$ of the $224$ policy canvas after resize; locations are redrawn uniformly per episode and held fixed within the episode.  

\paragraph{OpenVLA and UniVLA (single view).}
Victims observe only \texttt{agentview}.  The environment renders at $256{\times}256$; observations are resized to the policy resolution ($224$) and
the patch is applied afterwards.  
When the patch was trained as a PaliGemma dual-view specialist, we transfer only the \texttt{primary} file - wrist specialists are rejected by the single-view eval entrypoints.  UniVLA uses the same paste geometry and episode budget; the policy load path and action head differ, but the adversarial observation pipeline (area fraction, post-resize paste, random location) is shared.
\paragraph{$\pi_{0.5}$ (dual view).}
$\pi_{0.5}$ receives both cameras.  For PaliGemma dual-specialist attacks we paste \texttt{PRIMARY} on agentview and \texttt{WRIST} on eye-in-hand with independent uniform draws (sharing one specialist on both cameras is disallowed).  For a Prismatic single-view patch transferred to $\pi_{0.5}$, the same file may be applied
to both streams only under an explicit shared-patch override.  Policy inputs use
\texttt{resize\_with\_pad} to $224$, replan every $5$ steps, and the same $500$-episode budget and horizon table as above.


\section{More Details about Our Method}
\label{app:additional_method}

\subsection{Why $\mathcal{L}_\text{VIS}$ attack post-projector visual tokens?}
\label{app:post_projector}

Our threat model optimizes a universal image patch against the frozen ancestor vision--language model (VLM), rather than against the downstream robot policy. For the Prismatic ancestor, the vision backbone contains a DINOv2 tower and a SigLIP tower.  Their spatially corresponding features are concatenated along the channel dimension and then mapped by a frozen projector into the Llama embedding space:
\begin{equation}
    \mathbf{V}(x) =     \left[
        f_{\mathrm{DINO}}(x);
        f_{\mathrm{SigLIP}}(x)
    \right]
    \in \mathbb{R}^{N\times(d_D+d_S)},
    \qquad
    \mathbf{Z}(x)=g_{\mathrm{proj}}(\mathbf{V}(x))
    \in \mathbb{R}^{N\times d_{\mathrm{LM}}}.
    \label{eq:projector_interface}
\end{equation}
For the 224-pixel checkpoint used in our experiments, $N=16^2=256$ after removing non-spatial prefix tokens.  Importantly, the two towers are fused along the feature dimension; they are not appended as two token sequences.

We disrupt $\mathbf{Z}$ for two reasons.  First, it is the actual visual interface consumed by the language decoder and inherited by OpenVLA, so it captures distortions that survive both tower fusion and the visual-to-language alignment map.  Second, independently attacking the two pre-projector spaces can spend optimization capacity on directions removed or attenuated by $g_{\mathrm{proj}}$.  Post-projector disruption therefore targets a representation that is closer to the downstream policy's decision path.

\begin{algorithm}[t]
\caption{Factorized Caption$_0$ generation and selection}
\label{alg:Caption$_0$}
\begin{algorithmic}[1]
\Require frame $x$, instruction $t$, candidate count $K$
\Ensure $\text{Caption}_0=(c^{\mathrm{obj}},c^{\mathrm{attr}},c^{\mathrm{rel}})$ \Comment{the factorized side labels of Section~4.2}
\State $c^{\mathrm{obj}} \gets \Call{ParseTaskObject}{t}$
\If{$c^{\mathrm{obj}}$ is unresolved}
    \State $c^{\mathrm{obj}} \gets \Call{RecoverVisibleObject}{x,t}$
\EndIf
\State $\mathcal{A},\mathcal{R} \gets \Call{GenerateCandidates}{x,t,c^{\mathrm{obj}},K}$
    \Comment{frozen ancestor VLM; attribute and relation branches}
\For{$q \in \mathcal{A}\cup\mathcal{R}$}
    \State $q \gets \Call{NormalizePhrase}{q}$
    \If{$q$ is empty, truncated, action-bearing, leaked, implausible, or uses an invalid landmark}
        \State remove $q$ from its candidate pool
    \EndIf
\EndFor
\State extract instruction landmarks and multi-object constraints from $t$
\For{$q \in \mathcal{A}\cup\mathcal{R}$}
    \State compute within-frame $z_{\mathrm{CLIP}}(x,q)$ and $z_{\mathrm{SigLIP}}(x,q)$
    \State $S(q) \gets z_{\mathrm{SigLIP}}(x,q)+0.6\,z_{\mathrm{CLIP}}(x,q)$
    \If{$q\in\mathcal{R}$}
        \State $S(q) \mathrel{+}=0.8\,\mathrm{LandmarkShare}(q,t)$
        \State $S(q) \mathrel{+}=1.2\,\mathrm{ObjectCoverage}(q,t)$
        \State $S(q) \mathrel{-}=0.8\,\mathrm{PositionConflict}(q,t)$
    \EndIf
\EndFor
\State $c^{\mathrm{attr}}\gets\arg\max_{q\in\mathcal{A}}S(q)$
\State $c^{\mathrm{rel}}\gets\arg\max_{q\in\mathcal{R}}S(q)$
\State store aligned $(x,t,c^{\mathrm{obj}},c^{\mathrm{attr}},c^{\mathrm{rel}})$ and negatives
\State \Return $(c^{\mathrm{obj}},c^{\mathrm{attr}},c^{\mathrm{rel}})$
\end{algorithmic}
\end{algorithm}

\subsection{$\mathcal{L}_\text{IGS}$: Probe Construction and Evidence Computation}\label{app:probe and evidence}
This section details how the probe set $Q_b$ of \S\ref{sec:objective2} is constructed and how the evidence scores are computed from it.

\textbf{Extracting the factorized labels (Caption$_0$) and constructing probe set $\mathcal{Q}_b$.}
Caption$_0$ is our implementation name for the factorized side labels (object, attribute, relation) introduced in Section 4.2. It is generated offline by querying the frozen Prismatic VLM. To keep the language head from trivially recovering the probe word, the instruction is wrapped as ``What action should the robot take to \{instruction\}?", with known object and color words replaced by the placeholder ``object". 
The object field is recovered from the frame with the task instruction as context. Multiple attribute and relation candidates are then generated with separate prompts. The filtering stage rejects dirty text, prompt leakage, implausible foreign objects, relation fragments with truncated endings, and obvious attribute--relation role violations. It also cleans candidate text and reranks the retained pool using within-sample normalized CLIP and SigLIP image-text scores:
\begin{equation}
    S(q) = z_\mathrm{SigLIP}(x,q) + 0.6\cdot z_\mathrm{CLIP}(x,q) + R_\mathrm{structure}(q)
	\label{eq:caption_rerank}
\end{equation}
where $z(\cdot,\cdot)$ are within-frame normalized image-text scores and
$R_{\mathrm{structure}}$ rewards instruction-landmark sharing and multi-object coverage while penalizing spatial conflicts. The best candidate of each field forms the probe set $Q_b$. If a field is unavailable, we fall back to the corresponding noun parsed from the instruction. CLIP and SigLIP are used offline as labeling tools only; the attack itself never queries them. Algorithm. \ref{alg:Caption$_0$} gives the complete procedure.

\begin{algorithm}[t]
\caption{Staged visual-semantic patch optimization}
\label{alg:staged_patch}
\begin{algorithmic}[1]
\Require Frozen ancestor VLM $F_{\mathrm{VLM}}$; training frames and Caption$_0$ fields;
patch $\delta$; semantic start $K_1$; basin threshold $\gamma$
\State Initialize Adam state for $\delta$ and cosine EMA $c_t$
\For{$k=1,\ldots,T$}
    \State Sample a frame batch and cache detached clean visual features
    \If{$k> K_1$}
        \State Cache detached clean decoder probes and spatial weights
    \EndIf
    \State Set the cosine-scheduled learning rate for outer step $k$
    \For{$r=1,\ldots,4$} \Comment{four Adam updates per outer step}
        \State Zero the patch gradient
        \For{$s=1,\ldots,8$} \Comment{eight EOT draws per update}
            \State Transform and paste $\delta$; apply photometric jitter
            \State Compute patched post-projector tokens
            \State Compute $\mathcal{L}_{\mathrm{VIS}}$ and update $c_t$
            \If{$k> K_1$ \textbf{and} $c_t\leq\gamma$}
                \State Compute image-token logit-lens $\mathcal{L}_{\mathrm{IGS}}$
            \Else
                \State Set $\mathcal{L}_{\mathrm{IGS}}\gets 0$
            \EndIf
            \State Accumulate $\nabla_\delta(\mathcal{L}_{\mathrm{VIS}}+\lambda_{\mathrm{sem}}\,\mathcal{L}_{\mathrm{IGS}})/8$
        \EndFor
        \State Clip the patch-gradient norm; take one Adam step
        \State Project $\delta$ elementwise onto $[0,1]$
    \EndFor
\EndFor
\State \Return best-scoring and final patches
\end{algorithmic}
\end{algorithm}


\textbf{Training set.}  We optimize on the first agent-view frame (t = 0) of each demonstration. Each stored image is a $1\times3\times224\times224$ floating-point tensor normalized with ImageNet mean $(0.485,0.456,0.406)$ and standard deviation $(0.229,0.224,0.225)$.   For example, the LIBERO-Spatial setup contains 500 such samples. For BridgeData~V2, we use the fixed external camera (\texttt{image\_0} in the TFDS release; the analogue of LIBERO's agent view) and the same first-frame rule. From the Bridge train split we draw a stratified 6{,}000-frame subset over task families (\texttt{widowx\_*}, \texttt{other}, and unlabeled),with sampling seed 2026; empty-language episodes are retained. For the equal-size ablation in Sec.~5.2, we further draw 500 frames from this 6{,}000-frame pool without replacement.

\textbf{Frame preprocessing.} Stored frames are $224\times224$ RGB. LIBERO agent-view frames are rotated by $180^\circ$ to correct the environment's camera convention; BridgeData~V2 uses the fixed external camera without this correction. At optimization time the frame is resized to the attack canvas with antialiased bicubic interpolation, and the ancestors' own image transforms are then applied.

\textbf{Evidence computation.} We hook the output of decoder block $12$ (zero-indexed), retain the image-token states only, apply the frozen final RMSNorm, and score each probe word by its inner product with the corresponding rows of the frozen language-model head, averaged over subword   tokens and then over the probe set. The $256$ image tokens are mapped back to the native $16\times16$ grid, and the clean-image evidence defines a detached soft spatial weighting over all cells. The loss is the two-term softplus objective as \eqref{eq:IGS}, with $B=4$, $m=5$, $\tau_{\mathrm{drop}}=0.5$.

\textbf{Why factorized labels, not a single caption.} We use factorized labels instead of a single free-form caption for two reasons. First, the instruction identifies the task but does not uniquely describe the frame's current visual state; Caption$_0$ supplies frame-conditioned anchors, namely the manipulated object, its observed attributes, and a visible landmark relation. Second, factorization prevents a long caption from collapsing all supervision into one sequence-level likelihood: object, attribute, and relation evidence can be enabled and weighted independently. The objective never trains the VLM to emit these captions; Caption$_0$ only selects a small set of frozen language-head rows and clean-image spatial weights. When Caption$_0$ is absent, the loader falls back to nouns parsed from the instruction.

\textbf{Relation to the instruction.} Prismatic places the image tokens before the text prompt, so under causal attention the hooked image-token states cannot attend to the instruction. ``Instruction-grounded'' therefore describes how the probe set is chosen - the instruction and
the frame description determine \emph{which} vocabulary rows are probed.

\begin{figure}[t]
	\centering
	\includegraphics[width=0.55\linewidth]
	{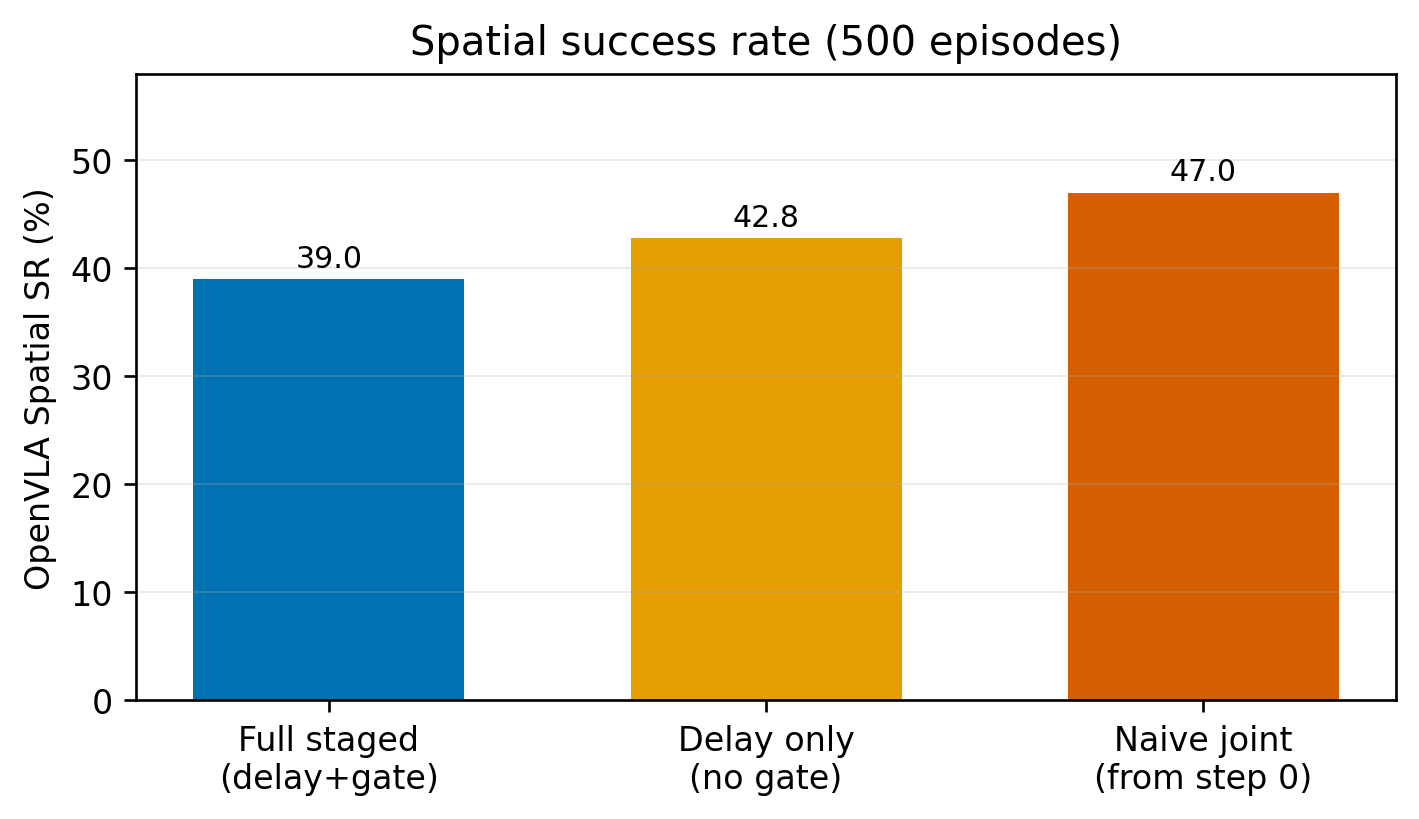}
	\caption{OpenVLA--LIBERO-Spatial success rates (500 episodes) under three joint recipes: full staging with a 300-step semantic delay and the visual-basin gate, the same delay without the gate, and a naive joint that enables the semantic objective from the beginning. }
	\label{fig:staging-sr}
\end{figure}

\begin{figure}[t]
	\centering
	\includegraphics[width=0.95\linewidth]
	{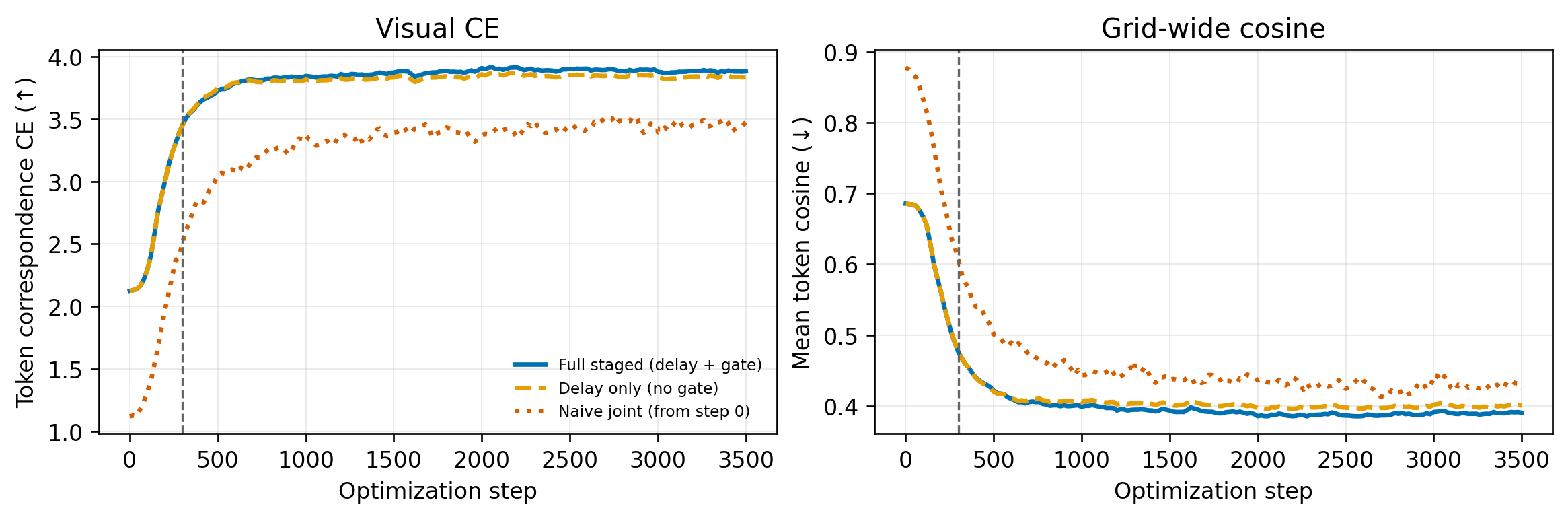}
	\caption{Optimization dynamics of the joint recipe on OpenVLA–LIBERO-Spatial. }
	\label{fig:joint_dynamics}
\end{figure}

\subsection{Why Staged Joint Optimization for $\mathcal{L}_{\mathrm{joint}}$?}
\label{app:staged}

Algorithm. \ref{alg:staged_patch} shows the staged joint optimization process for $\mathcal{L}_{\mathrm{joint}}  
=\mathcal{L}_{\mathrm{VIS}}
+\beta_k\,\lambda_{\mathrm{sem}}\,\mathcal{L}_{\mathrm{IGS}}$, where $\beta_k=\mathbb{I}(k > K_1)\cdot\mathbb{I}(c_t\le \gamma)$. 
We set $\lambda_{\mathrm{sem}}=0.5$, $K_1=300$, $\gamma=0.45$. The warmup stabilizes the optimizer, whereas the delay gives the visual term time to break token correspondence and lower the grid-wide cosine before semantic suppression is admitted. 

Here we clarify the schedule and show that both parts are necessary. Figure ~\ref{fig:staging-sr} shows the ablation on the LIBERO-Spatial suite: the full staged recipe reaches 39.0\% SR, removing only the basin gate raises it to 42.8\%, and enabling the semantic objective from step 0 (naive joint) raises it to 47.0\%, so removing either constraint weakens the attack. 
Figure \ref{fig:joint_dynamics} shows the training dynamics, tracking the two quantities optimized by $\mathcal{L}_{\mathrm{VIS}}$: the grid-wide cosine similarity (the absolute term $\mathcal{L}_{\mathrm{abs}}^v$) and the correspondence cross-entropy (the relative term $-\mathcal{L}_{\mathrm{rel}}^v$):
after semantic activation, the staged recipe keeps a slightly lower grid-wide cosine while maintaining a comparable correspondence CE, whereas the ungated run achieves a larger semantic-evidence drop that does not translate into a stronger downstream attack. Maximizing the semantic diagnostic alone is therefore insufficient without first stabilizing the visual basin.

\end{document}